\documentclass[amstex,12pt]{article}

\usepackage{graphicx}
\usepackage{amsfonts}
\usepackage{amssymb}
\usepackage{amsmath}

\newtheorem{thm}{Theorem}[section]
\newtheorem{lm}[thm]{Lemma}
\newtheorem{prop}[thm]{Proposition}

\newtheorem{q}[thm]{Question}
\newtheorem{rmk}[thm]{Remark}
\newtheorem{ex}[thm]{Example}
\newtheorem{df}[thm]{Definition}

\begin{document}

\author{Jin-ichi Itoh and Costin V\^\i lcu}
\title{Cut locus structures on graphs}
\maketitle

\noindent {\bf Abstract.} {\small
Motivated by a fundamental geometrical object, the cut locus, we introduce and study a new combinatorial structure on graphs.
\\{\bf Math. Subj. Classification (2000):} 57R70, 53C20, 05C10}


\section{Introduction}

The motivation of this work comes from a basic notion in riemannian geometry, that we shortly present in the following.
In this paper, by a {\it surface} we always mean a complete, compact and connected 
$2$-dimensional riemannian manifold without boundary.

The {\it cut locus $C(x)$ of} the point $x$ in the surface $S$ is the set of all extremities (different from $x$) 
of maximal (with respect to inclusion) shortest paths (geodesic segments) starting at $x$; 
for basic properties and equivalent definitions refer, for example, to \cite{ko} or \cite{sa}.
The notion was introduced by H. Poincar\'e \cite{p} and gain, since then, an important place in global riemannian geometry.

For surfaces $S$ is known that $C(x)$, if not a single point, is a local tree 
(i.e., each of its points $z$ has a neighbourhood $V$ in $S$ such that 
the component $K_z(V)$ of $z$ in $C(x)\cap V$ is a tree), 
even a tree if $S$ is homeomorphic to the sphere.
A {\it tree} is a set $T$ any two points of which can be joined
by a unique Jordan arc included in $T$.

{\sl All our graphs are finite, connected, undirected, and may have multiple edges or loops}.

S. B. Myers \cite{M} established that the cut locus of a real analytic surface is (homeomorphic to) a graph, 
and M. Buchner \cite{Bu2} extended the result for manifolds of arbitrary dimension.
For not analytic riemannian metrics on $S$, cut loci may be quite large sets, see the work of 
J. Hebda \cite{H1} and of the first author \cite{I2}.
Other contributions to the study of this notion were brought, among others, by 
M. Buchner \cite{Bu1}, \cite{Bu3}, H. Gluck and D. Singer \cite{GS}, J. Hebda \cite{H2}, 
J. Itoh \cite{I1}, K. Shiohama and M. Tanaka \cite{ST}, T. Zamfirescu \cite{z-ep}, \cite{z-cl}, A. D. Weinstein \cite{W}.

We show in another paper \cite{iv2} that 
{\sl for every graph $G$ there exists a surface $S_G$ and a point $x$ in $S$ whose cut locus 
$C(x)$ is isomorphic to $G$}; rephrasing, {\sl every graph can be realized as a cut locus}.

If $G$ has an odd number $q$ of generating cycles then any surface $S_G$ realizing $G$ is non-orientable,
but if $q$ is even then one cannot generally distinguish, by simply looking to the graph $G$, 
whether $S_G$ is orientable or not: explicit examples show that both possibilities can occur \cite{iv3}.
In other words, seen as a graph, {\sl the cut locus does not encode the orientability of the ambient space}. 

This is our main motivation to endow graphs with a combinatorial structure -- 
that of cut locus structure, or shortly CL-structure.

In this paper we treat combinatorial aspects of this new notion:
in Section \ref{def} we introduce and discuss this notion, 
in Section \ref{repres} we give two planar representations of CL-structures,
and in the last section we enumerate all such structures on ``small'' graphs.

In a second paper \cite{iv2} we show that every CL-structure actually corresponds to a cut locus on a surface, 
while in a subsequent one \cite{iv3} we consider the orientability of the surfaces realizing CL-structures as cut loci. 
In par\-ti\-cu\-lar, any graph endowed with a CL-structure does encode the orientability of the ambient space where is lives as a cut locus.
An upper bound on the number of CL-structures on a graph is given in \cite{iv4}.

\medskip

At the end of this section we recall a few notions from graph theory, in order to fix the notation.

Let $G$ be a graph with vertex set $V=V(G)$ and edge set $E=E(G)$.
Denote by $B$ the set of all {\it bridges} in the graph $G$; i.e., edges whose removal disconnects $G$. 
Each non-vertex component of $G \setminus B$ is called a $2$-{\it connected component} of $G$.

A $k${\it -graph} is a graph all vertices of which have degree $k$.

The power set ${\cal E}$ of $E$ becomes a $Z_2$-vector space over the two-element field $Z_2$ if endowed with 
the symmetric difference as addition. 
${\cal E}$ can be thought of as the space of all functions $E \to Z_2$, 
and called the (binary) {\it edge space} of $G$. 
The (binary) {\it cycle space} is the subspace ${\cal Q}$ of ${\cal E}$ generated by 
(the edge sets of) all simple cycles of $G$.
If $G$ is seen as a simplicial complex, ${\cal Q}$ is the space of $1$-cycles of $G$ with mod $2$ coefficients.


\section{Cut locus structures}
\label{def}

\begin{df}
A $G$-{\sl patch} on the graph $G$ is a topological surface $P_G$ with boundary, containing (a graph isomorphic to) $G$ and contractible to it.
\end{df}

\begin{rmk}
Every boundary component of a patch is homeomorphic to a circle, as a $1$-dimensional manifold without boundary.
\end{rmk}

\begin{df}
A $G$-{\sl strip} (or a strip on $G$, or simply a {\sl strip}, if the graph is clear from the context), 
is a $G$-patch with $1$-component boundary; i.e., whose boundary is one topological circle; 
see Figure 1 (a).
\end{df}

The next remark gives the geometrical background for the notion of cut locus structure.

\begin{rmk}
\label{constr_clns}
Consider a point $x$ on a surface $S$, and a geodesic segment $\gamma: [0,l] \to S$ parameterized by arclength, 
with $\gamma(0)=x$ and $\gamma(l) \in C(x)$.
For $\varepsilon >0$ smaller than the injectivity radius at $x$, and hence smaller than $l$, 
the point $\gamma(l-\varepsilon)$ is well defined.
Since $S \setminus C(x)$ is contractible to $x$ along geodesic segments, and thus homeomorphic to an open disk, 
the union over all $\gamma$s of those points $\gamma(l-\varepsilon)$ is homeomorphic to the unit circle, and therefore
the set $\bigcup_{\gamma} \{\gamma(l-\mu) : 0 \leq \mu \leq \varepsilon\}$ is a $C(x)$-strip.
\end{rmk}

\begin{df}
A {\sl cut locus structure} (shortly, a {\sl CL-structure}) on the graph $G$ is a strip on the cyclic part $G^{cp}$ of $G$.
\end{df}

\begin{rmk}
We show in another paper {\rm \cite{iv2}}, with geometrical tools, the converse to Remark \ref{constr_clns}: 
every CL-structure can be obtained (with some suitable surface and point on the surface) 
as described in Remark \ref{constr_clns}.
\end{rmk}

\begin{figure}
\centering
  \includegraphics[width=0.6\textwidth]{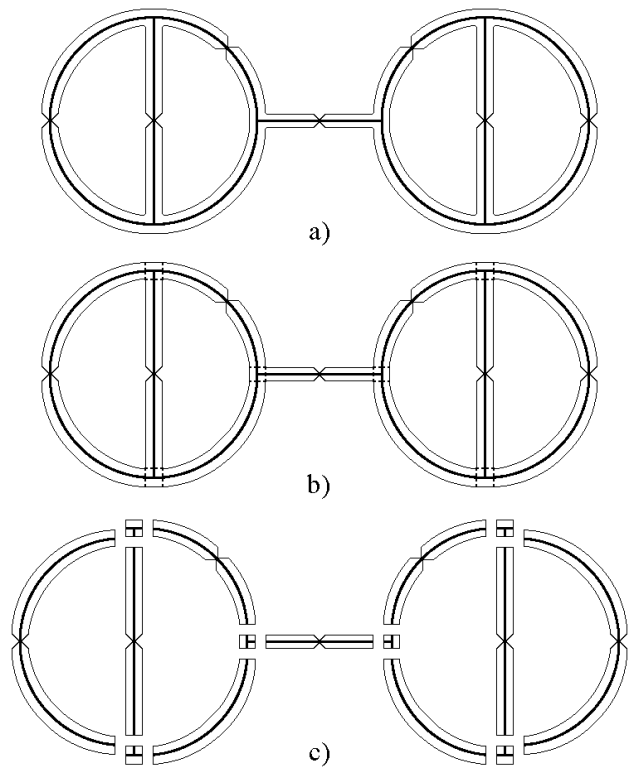}
\caption{A strip and its elementary decomposition.}
\label{Elem_decomp}
\end{figure}

\begin{rmk}
Each $G$-strip defines a circular order around each vertex of $G$, and thus a {\sl rotation system}.
Conversely, one can alternatively define a $G$-strip as the graph associated to a rotation system, 
together with a 2-cell embedding having precisely one face.
We choose not to follow this way, and to keep in our presentation as much as possible of the geometrical intuition.
\end{rmk}

\begin{df}
An {\sl elementary strip} is an edge-strip (arc-strip) or a point-strip; i.e., a strip defined by the graph with precisely one edge (arc) of different extremities, respectively by the graph consisting of one single vertex.
\end{df}

\begin{df}
An {\sl elementary decomposition} of a $G$-patch $P_G$ is a decomposition of $P_G$ into elementary strips such that:
\\- each edge-strip corresponds to precisely one edge of $G$;
\\- each point-strip corresponds to precisely one vertex of $G$;
see Figure \ref{Elem_decomp} (b) and (c).
\end{df}

\begin{rmk}
Our notion of ``$G$-patch'' is equivalent to that of ``{\sl fibered surface}'' introduced by M. Bestvina and M. Handel: 
``{\sl a fibered surface is a compact surface $F$ with boundary which is decomposed into arcs and into polygons 
that are modeled on $k$-junctions, $k=1,2,3,...$. The components of the subsurface fibered by arcs are strips. 
Shrinking the decomposition elements to points produces a graph $G$, where vertices (of valence $k$) correspond to ($k$-) junctions and strips to edges. We can think of $G$ as being embedded in $F$, representing the spine of $F$}'' {\rm \cite{Best-Han}}.
We choose the most (in our opinion) appropriate name for our purpose, and thus different from theirs.
\end{rmk}

In order to easier handle a CL-structure, we associate to it an object of combinatorial nature.
To this goal, denote by ${\cal P}$ and ${\cal A}$ the set of all point-strips, respectively edge-strips, 
of a CL-structure ${\cal C}$ on the graph $G$.

\begin{df}
Consider an elementary decomposition of the $G$-strip $P_G$ such that each elementary strip has a distinguished face, 
labeled $\bar 0$. The face opposite to the distinguished face will be labeled $\bar 1$. 
Here, $\bar 0$ and $\bar 1$ are the elements of the $2$-element group $(Z_2, \oplus)$. 

To each pair $(v,e) \in V \times E$ consisting of a vertex $v$ and an edge $e$ incident to $v$, 
we associate the $Z_2$-sum $\bar s (v,e)$ of the labels of the elementary strips $\nu \in {\cal P}$, 
$\varepsilon \in {\cal A}$ associated to $v$ and $e$; i.e.,
$\bar s (v,e) = \bar 0$ if the distinguished faces of $\nu$ and $\varepsilon$ agree to each other, and $\bar 1$ otherwise.
Therefore, to any cut locus structure ${\cal C}$ we can associate a function $s_{\cal C} :E \to \{\bar 0,\bar 1\}$, 
\begin{eqnarray}
\label{companion}
s_{\cal C}(e)= \bar s (v,e) \oplus \bar s (v',e),
\end{eqnarray}
where $v$ and $v'$ are the vertices of the edge $e \in E$.

We call the function $s_{\cal C}$ defined by (\ref{companion}) the {\sl companion function} of ${\cal C}$. 
\end{df}

The value $s_{\cal C} (e)$ above can be thought of as the {\sl switch} of the edge $e$.

\begin{df}
\label{diff_CL}
Assume first that the graph $G$ is $2$-connected.
Two {\sl CL-structures} ${\cal C}$, ${\cal C}'$ on $G$ are called {\sl equivalent} if 
their {\sl companion functions} are {\sl equivalent}: i.e., 
$s_{\cal C}$ and $s_{{\cal C}'}$ are equal, up to a simultaneous change of 
the distinguished face for all elementary strips in $G$
(i.e., either $s_{\cal C} = s_{{\cal C}'}$, or $s_{\cal C} =\bar 1 \oplus s_{{\cal C}'}$).

If $G$ is not $2$-connected, the {\sl CL-structures} ${\cal C}$, ${\cal C}'$ on $G$ are called {\sl equivalent} if their companion functions are equivalent on every $2$-connected component of $G$. See Figure 2.
\end{df}

\begin{figure}
\label{Equivalent CL-structures}
\centering
  \includegraphics[width=0.45\textwidth]{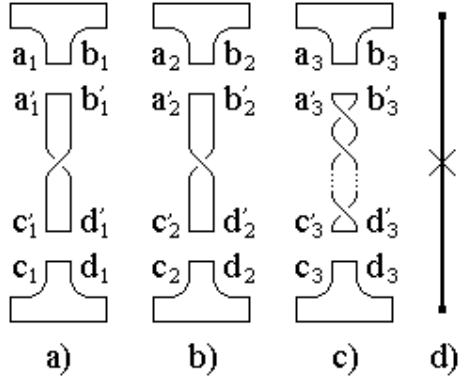}
\caption{Equivalent CL-structures (a), (b) and (c), and schematic representation (d).
The edge-strip at (a) corresponds to a rectangular band whose base is $\pi$-rotated ``to the left'' with respect to the top;
the edge-strip at (b) corresponds to a rectangular band whose base is $\pi$-rotated ``to the right'' with respect to the top;
the edge-strip at (c) corresponds to a rectangular band whose base is $(2k+1)\pi$-rotated ``to the left'' 
with respect to the top.}
\end{figure}

\begin{df}
\label{def_switched_edge}
An edge-strip $P_e$ (or simply an edge $e$) in a CL-structure ${\cal C}$ is called {\sl switched} if $s_{\cal C} (e)=\bar 1$.
\end{df}

\begin{prop}
If two CL-structures on the same graph $G$ are equivalent then the corresponding $G$-strips are homeomorphic surfaces.
\end{prop}

\noindent{\sl Proof:} 
We may assume that $G$ is cyclic.

Assume, moreover, that we have two CL-structures on $G$, whose companion functions are equivalent 
on every $2$-connected component of $G$.
The desired homeomorphism can be constructed inductively, extending it with each new ``gluing'' of an elementary strip, 
see Figure \ref{Equivalent CL-structures}.
\hfill $\Box$


\section{Representations of CL-structures}
\label{repres}

We propose two ways to planary represent a CL-structure ${\cal C}$ on the graph $G$.

\begin{df}
The {\sl graph representation} of ${\cal C}$ starts with some planar representation of $G$, and afterward points out the CL-structure, see Figure 3 (a).
\end{df}

\begin{figure}
\label{Equiv_represent}
\centering
  \includegraphics[width=0.75\textwidth]{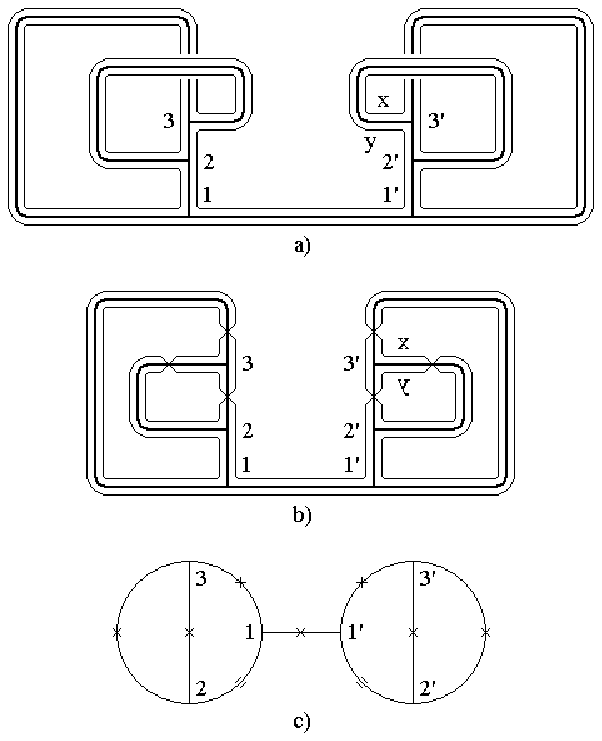}
\caption{Representations of CL-structures. a) {\it Graph representation} of a strip. b) Intermediate step to obtain (c). 
c) {\it Natural representation} for the strip at (a). Additional points $x,y$ are indicated to make clear the transformation.}
\end{figure}

\begin{df}
The {\sl natural representation} of ${\cal C}$ starts by representing in the plane each vertex-strip such that 
its distinguished face is ``up'', and afterward connects the vertex-strips by edge-strips.
The idea is illustrated by Figure \ref{Equiv_represent} (b) and (c).
\end{df}

\begin{figure}
\label{Schem_repres}
\centering
  \includegraphics[width=0.6\textwidth]{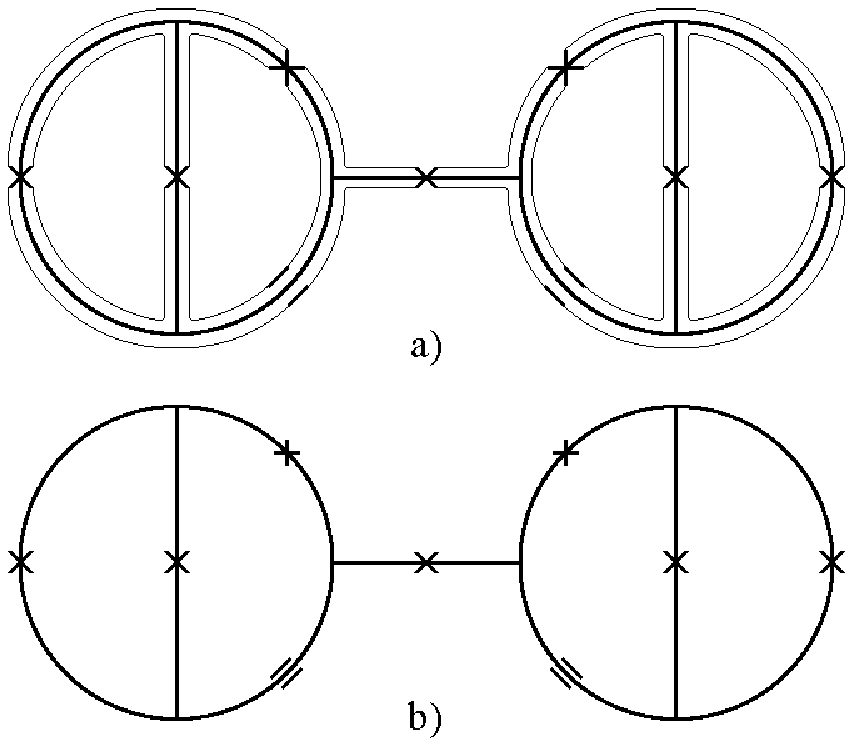}
\caption{Schematic representation of the strip in Figure \ref{Elem_decomp} (a).}
\end{figure}

\begin{rmk}
\label{x=}
Consider the natural representation of a CL-structure on a cubic graph.
We shall overwrite an ``{\rm x}'' to the drawn image of an edge if its strip is switched, 
and an ``{\rm =}'' to the drawn image of an edge if its strip is not-switched. 
See Figures 4 and 3.
\end{rmk}

\begin{rmk}
Neither the natural representation, nor the graph representation, of a CL-structure on a graph is unique.
\end{rmk}

\begin{prop}
\label{equiv_repres}
For any planar cubic graph $G$ and any CL-structure on $G$, the natural representation and the graph representation coincide,
up to planar homeomorphisms.
\end{prop}

\noindent{\sl Proof:} This follows from the definitions above.
\hfill $\Box$

\begin{figure}
\label{Flat_torus_CLNS}
\centering
  \includegraphics[width=0.6\textwidth]{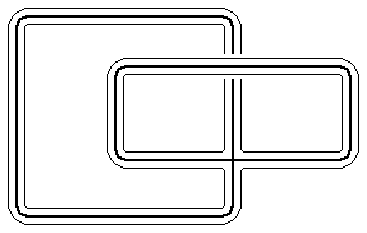}
\caption{CL-structure obtained from a flat torus of rectangular fundamental domain.}
\end{figure}

\begin{ex}
\label{planar_graph_crossings}
If the $3$-graph $G$ is not planar, Proposition \ref{equiv_repres} is not true.
An easy example, obtained from a flat torus of rectangular fundamental domain 
(see the procedure described in Remark \ref{constr_clns}), is illustrated by Figure 5.
\end{ex}

Directly from the definitions we have the following.

\begin{lm}
\label{cy-sw}
In any natural representation of a strip, each cycle-patch contains at least one switched edge-strip.
\end{lm}

We can give four open questions.

\begin{q}
Characterize the companion functions of CL-structures in the set 
${\cal S}=\{ s:E \to \{\bar 0, \bar 1\} \}$.
\end{q}

\begin{q}
A planar graph is, by definition, a graph which can be represented in the plane without crossings (self-intersections). 
As we have seen in Example \ref{planar_graph_crossings}, 
there are CL-structures on (not cubic) planar graphs whose natural representasions in the plane 
necessarily produce crossings. What is the minimal number of such crossings which guarantees a planar natural representation?

The same question can be asked for non planar graphs too, 
where the (minimal number of necessary) crossings of the graphs is a new parameter.
\end{q}

\begin{q}
How many CL-structures can coexist on the same graph?
\end{q}

Some (not sharp) upper bound will be given in \cite{iv3}.

\begin{q}
Which of the graphs with $q$ generating cycles has the largest number of different CL-structures?
\end{q}

We shall address in the following section the last two questions above, for graphs with two and three generating cycles.


\section{CL-structures on small graphs}
\label{small_graphs}


We present in this section all distinct cut locus structures on $3$-graphs with $q=2,3$ generating cycles.

The following statement can be obtained by straightforward inductive constructions.

\begin{figure}
\label{Fig_2cy}
\centering
  \includegraphics[width=0.8\textwidth]{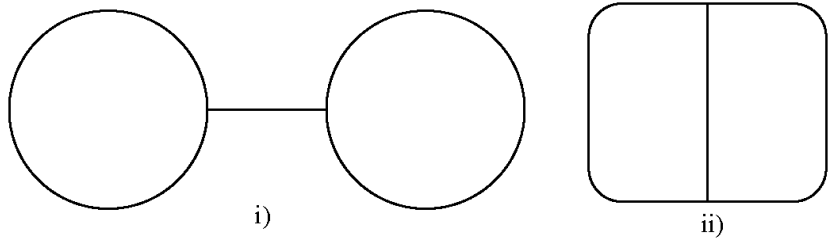}
\caption{All $3$-graphs with $2$ generating cycles.}
\end{figure}

\begin{lm}
\label{g43}
There are precisely $2$, respectively $6$, distinct $3$-graphs with 
$2$, respectively $3$, generating cycles, see Figures 6 and 7.
\end{lm}

\begin{figure}
\label{Fig_3cy}
\centering
  \includegraphics[width=0.8\textwidth]{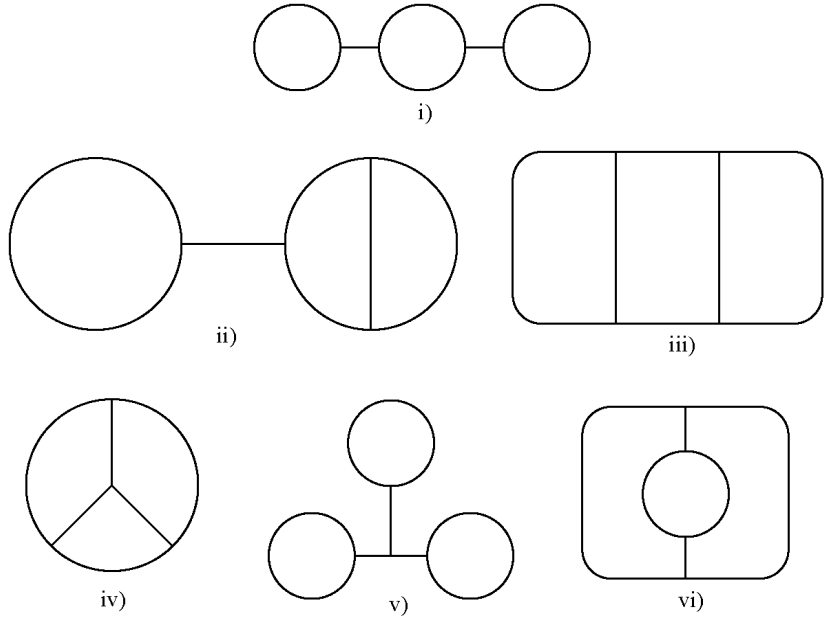}
\caption{All $3$-graphs with $3$ generating cycles.}
\end{figure}

\begin{thm}
\label{4}
a) There are precisely $3$ non-equivalent CL-structures on the $3$-graphs with $2$ generating cycles, see Figures 8 and 9.

b) There are precisely $17$ non-equivalent CL-structures on the $3$-graphs with $3$ generating cycles, see Figures 10 -- 15.
\end{thm}

\begin{figure}
\label{Fig_2cy_i}
\centering
  \includegraphics[width=0.5\textwidth]{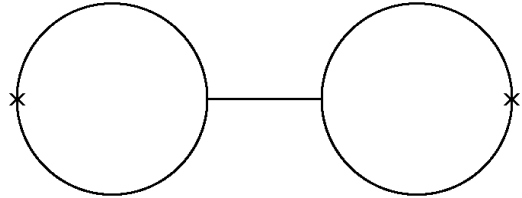}
\caption{Unique CL-structure on the graph in Figure 6 (i).}
\end{figure}

\begin{figure}
\label{Fig_2cy_ii}
\centering
  \includegraphics[width=0.6\textwidth]{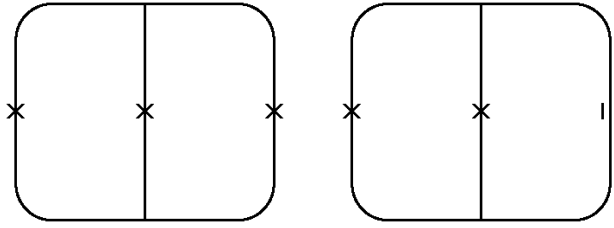}
\caption{$2$ CL-structures on the graph in Figure 6 (ii).}
\end{figure}

\noindent{\sl Proof:}
We employ the natural representation of CL-structures.
It is straightforward to generate all patches on the graphs in Figures 6 -- 7, 
to keep only the strips (by the use of Lemma \ref{cy-sw}), and 
to use Definition \ref{diff_CL} and the symmetries of the graphs to identify equivalent CL-structures.
\hfill $\Box$

\bigskip

\begin{figure}
\label{Fig_3cy_i}
\centering
  \includegraphics[width=0.5\textwidth]{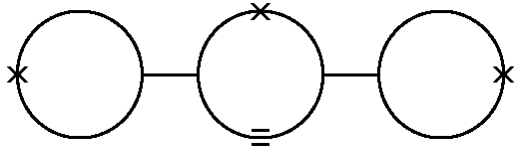}
\caption{Unique CL-structure on the graph in Figure 7 (i).}
\end{figure}

\begin{figure}
\label{Fig_3cy_ii}
\centering
  \includegraphics[width=0.85\textwidth]{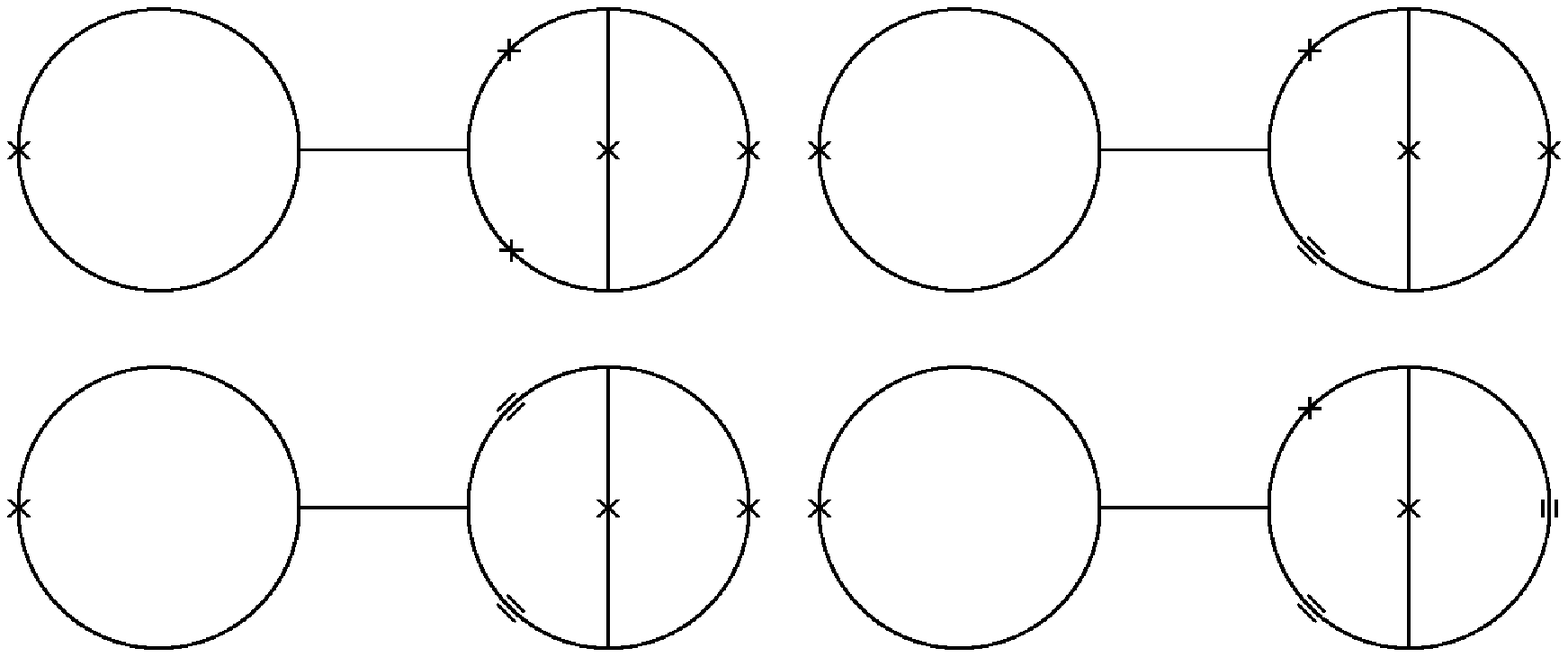}
\caption{$4$ CL-structures on the graph in Figure 7 (ii).}
\end{figure}

\begin{figure}
\label{Fig_3cy_iii}
\centering
  \includegraphics[width=0.65\textwidth]{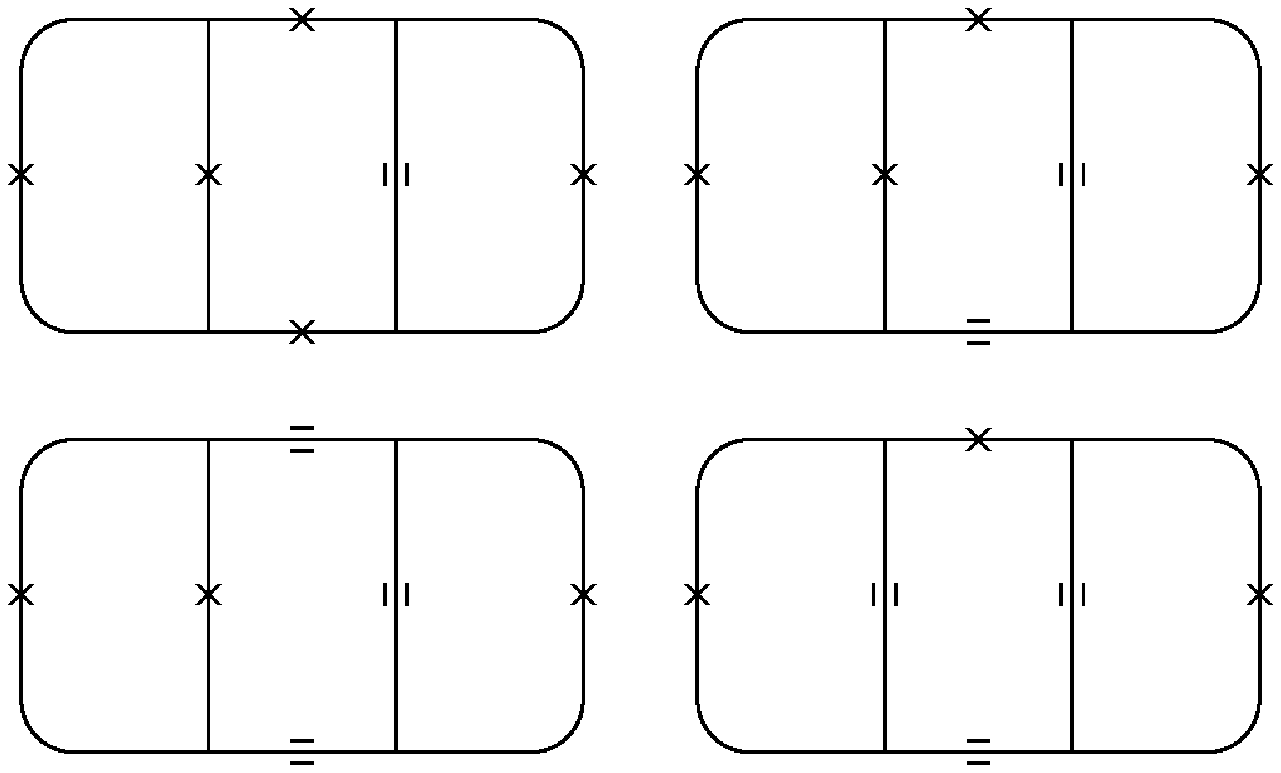}
\caption{$4$ CL-structures on the graph in Figure 7 (iii).}
\end{figure}

\begin{figure}
\label{Fig_3cy_iv}
\centering
  \includegraphics[width=0.65\textwidth]{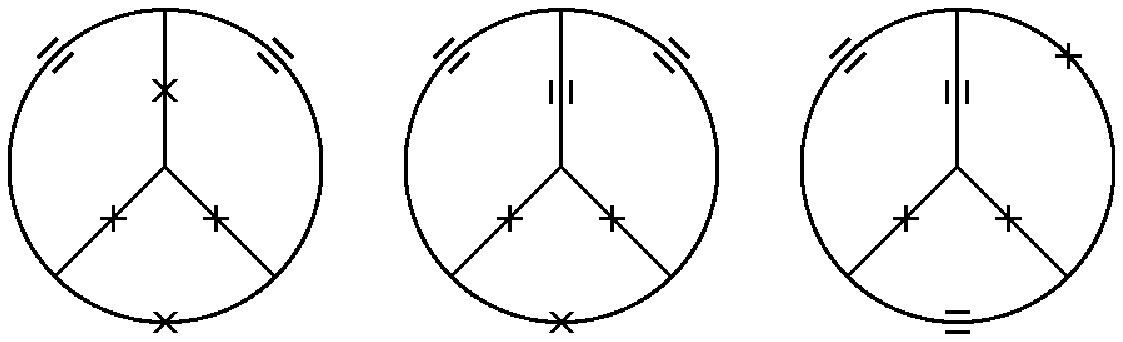}
\caption{$3$ CL-structures on the graph in Figure 7 (iv).}
\end{figure}

\begin{figure}
\label{Fig_3cy_v}
\centering
  \includegraphics[width=0.4\textwidth]{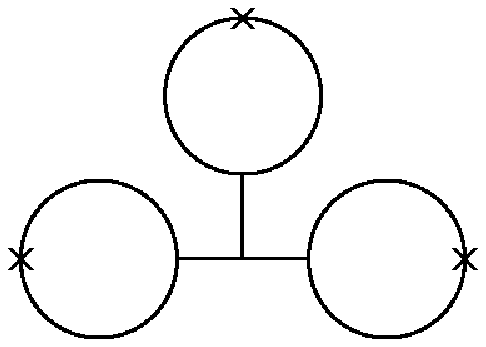}
\caption{Unique CL-structure on the graph in Figure 7 (v).}
\end{figure}

\begin{figure}
\label{Fig_3cy_vi}
\centering
  \includegraphics[width=0.6\textwidth]{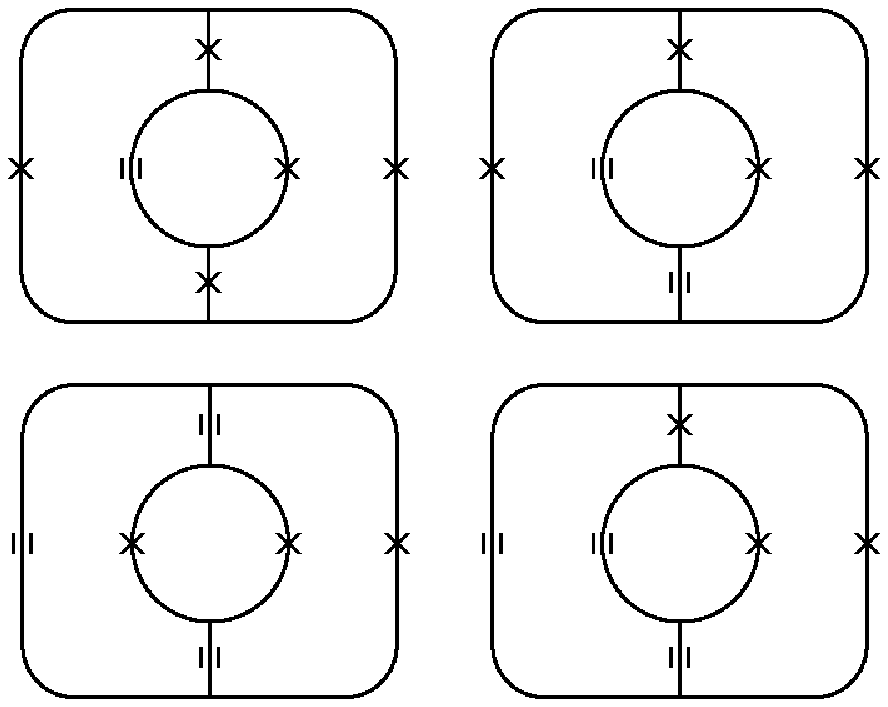}
\caption{$4$ CL-structures on the graph in Figure 7 (vi).}
\end{figure}

Our last result shows  that the case of CL-structures on $3$-graphs is, in some sense, sufficient.
For, define the {\it degree of a graph} as the maximal degree of its vertices.

\begin{thm}
\label{aprox}
Any CL-structures on a graph with $q$ generating cycles and degree larger than $3$ can be obtained from 
CL-structures on $3$-graphs with $q$ generating cycles, by contracting non-switched edge-strips.
\end{thm}

\noindent{\sl Proof:}
Fix $q$; we consider only graphs with $q$ generating cycles, and 
proceed by induction over the number of vertices of degree larger that $3$.
Denote by $D(G)$ this number for the graph $G$.

Assume the cyclic graph $G$ has $D(G) \geq 1$, and choose a vertex $v$ in $G$ with deg$(v)=d>3$.

Let ${\cal C}$ be a CL-structure on $G$, and denote by $v_1, ..., v_d$ the neighbours of $v$ in $G$, 
and by $T$ the subtree of $G$ rooted at $v$, with leaves $v_1, ..., v_d$.
Let $G^-$ be the complement of $T$ in $G$, and ${\cal C}^-$ be the union of patches naturally induced by ${\cal C}$ on $G^-$.
Let $s_{\cal C}^-$ be the restriction of the companion function $s_{\cal C}$ of ${\cal C}$ to $G^-$.

Replace $T$ in $G$ by a tree $T_3$ of leaves $v_1, ..., v_d$, all of whose internal vertices have degree 3
($T_3$ is generally not unique), and denote by $G^v$ the new graph.
Now complete ${\cal C}^-$ to a CL-structure ${\cal C}^v$ on $G^v$,
by extending $s_{\cal C}^-$ on the internal edges of $T_3$ with $\bar 0$, 
and on the external edges of $T_3$ with the original values of $s_{\cal C}$.
Observe that ${\cal C}^v$ is indeed a CL-structure on $G^v$, 
and $D(G^v)=D(G) -1$, so the proof is complete.
\hfill $\Box$

\bigskip

\noindent {\bf Acknowledgement } 
C. V\^\i lcu was partially supported by the grant PN II Idei 1187 of the Romanian Government.



Jin-ichi Itoh

\noindent {\small Faculty of Education, Kumamoto University
\\Kumamoto 860-8555, Japan
\\j-itoh@gpo.kumamoto-u.ac.jp}

\medskip

Costin V\^\i lcu

\noindent {\small Institute of Mathematics ``Simion Stoilow'' of the Romanian Academy
\\P.O. Box 1-764, Bucharest 014700, Romania
\\Costin.Vilcu@imar.ro}

\end{document}